
\lineskip=3pt minus 2pt
\lineskiplimit=3pt
\magnification=1200
\overfullrule 1pt
\baselineskip=20pt plus 2pt

{\bf\centerline {THE EVOLUTION OF NON-LINEAR SUB-HORIZON SCALE
}}
{\bf\centerline {ENTROPY FLUCTUATIONS IN THE EARLY UNIVERSE}}

\vskip 0.9in
\baselineskip=12pt plus 2pt
\centerline {{\sl Karsten Jedamzik} and {\sl George M. Fuller}}
\vskip 0.08in
\centerline {Department of Physics}
\centerline {University of California, San Diego}
\centerline {La Jolla, CA 92093-0319}
\vskip 1.85in
{\bf\centerline {ABSTRACT}}
\vskip 0.075in
\baselineskip=12pt plus 1pt
We examine the damping of non-linear sub-horizon scale entropy
fluctuations in early epochs of the universe
($T\approx 100$ GeV to $T\approx 1$ keV) by
neutrino, baryon, and photon induced dissipative processes.
Results of numerical evolution calculations are presented for
broad ranges of initial fluctuation amplitudes and length scales.
These calculations include a detailed treatment of neutrino
inflation, neutron and proton diffusion, photon diffusive heat
transport, and hydrodynamic expansion with photon-electron Thomson
drag. Neutrino inflation is treated both in the
diffusive heat transport regime where neutrinos are optically
thick on the length scales of the fluctuations, and in the
homogeneous heating regime where neutrinos are optically thin
on these scales.
We find considerable convergence in amplitude evolution for
appreciable ranges in initial fluctuation length scales and
amplitudes. Fluctuations produced with the right characteristics
at very early times ($T$ $^>_{\sim}$ $100$ GeV) are found to
survive through the nucleosynthesis epoch.

\vskip 0.97in
\centerline {\it Subject headings: cosmology: early universe -
hydrodynamics - radiative transfer}
\vfil\eject

\baselineskip=24pt plus 2pt

{\bf\centerline {1. Introduction}}
\vskip 0.1in

The present model for the history of the early universe envisions
significant departures from local thermal and chemical
equilibrium associated with symmetry breaking events.
These events might include, for example,
the QCD epoch ($T\approx 100$ MeV), the electroweak transition
($T\approx 100$ GeV), and an inflationary epoch.
Though the details and conclusions remain uncertain such
departures from thermal and chemical equilibrium at early epochs
have been proposed as production sites for entropy fluctuations.
By entropy fluctuations we mean any local deviations in the
entropy per baryon from the cosmic average value. Entropy
fluctuations could also result from processes associated with
primordial black holes,
cosmic strings, and domain walls. In this paper we wish to
examine how an entropy fluctuation produced with a given amplitude
and spatial dimension at a given epoch will subsequently evolve.
We will not be concerned here with how fluctuations are produced, but
rather with how their amplitudes and spatial scales change in time
due to the expansion of the universe and neutrino, baryon, and
photon dissipative processes.

The abundances of the elements emerging from the nucleosynthesis
epoch can be sensitive to the existence of entropy fluctuations
(cf. Mathews et al. 1990; Kurki-Suonio et al. 1988;
Terasawa \& Sato 1989abc, 1990; Jedamzik, Fuller, \& Mathews 1993;
and the review of nonstandard
Big Bang nucleosynthesis by Malaney \& Mathews 1993).
Observed light element abundances can be used to constrain
inhomogeneities at the nucleosynthesis epoch.
It is hoped that these limits, in turn,
may be used to constrain any processes
in the very early universe which might
generate entropy fluctuations.
Successfully being able to constrain the physics of the
early universe in this manner would
depend on (1) knowing how fluctuations evolve
from production at early
epochs through the nucleosynthesis epoch, and (2)
what the effects of such
fluctuations on nucleosynthesis would be.
This paper addresses the first
of these issues.

The production of entropy fluctuations on sub-horizon
scales in the non-linear
regime (amplitude greater than unity) has been discussed
extensively in the context of the QCD epoch
(cf. Witten 1984; Applegate \& Hogan 1985; Applegate, Hogan,
\& Scherrer 1987; Kajantie \& Kurki-Suonio 1986;
Kurki-Suonio 1988; Fuller, Mathews
\& Alcock 1988; Applegate 1991; Malaney \& Mathews 1993).
There is no consensus on fluctuation characteristics to be
expected from this epoch. In fact, recent lattice QCD results
(cf. Brown et al. 1990)
indicate that the chiral symmetry transition
associated with this epoch leads to a second order phase
transition and, therefore, no phase separation and no fluctuation
production.
However, these results are not definitive, as they are dependent
on lattice size and a proper treatment of dynamical quarks
(Petersson 1991).
Fluctuations associated with the QCD epoch might be generated
by kaon condensate nuggets (Nelson 1990) through a mechanism
which is independent of the order of the chiral symmetry phase
transition.

The electroweak phase transition is expected to be weakly first
order (Kirzhnits \& Linde 1976).
It has been proposed recently that baryogenesis may be associated
with non-equilibrium processes proceeding on the surfaces of
bubbles of the low temperature phase in the cosmic electroweak
transition
(Shaposhnikov 1986; 1987; 1988;
McLerran 1989; Cohen, Kaplan, \& Nelson 1990; 1991ab;
Turok \& Zadrozny 1990; 1991; Dine et al. 1991;
McLerran et al. 1991).
It may well be that the entropy-per-baryon and, hence, the
baryon-to-photon ratio resulting from these baryogenesis
scenarios, will be left with an inhomogeneous distribution
across the horizon (Fuller et al. 1993).
The generation of these entropy fluctuations is related to the
details of the non-equilibrium processes as well as the stochastic
nature of bubble nucleation and coalescence.

Other events or processes in the early universe have been
suggested as entropy fluctuation generators. For example
superconducting cosmic strings and associated large currents
and magnetic fields could generate appreciable inhomogeneity
in baryon density (Malaney \& Butler 1989). It is conceivable
that an inflationary epoch could generate non-linear small-scale
isocurvature fluctuations (Dolgov \& Silk 1992). For a
review of entropy fluctuation generation mechanism we refer
the reader to Malaney \& Mathews (1993).

Our goal in this paper is to describe the evolution in time
of the amplitude, size, and shape of a non-linear sub-horizon
scale fluctuation with properties specified at some initial
epoch. This problem has been discussed in previous work
(Peebles 1965; Misner 1967; Hogan 1978; Heckler \& Hogan 1993; Hogan 1993).
Our calculations differ from previous studies in that we extend our
survey to much higher temperatures ($T\approx 100$ GeV), we
perform complete numerical multi-spatial-zone calculations including
all relevant dissipative and hydrodynamic effects, we use
a more sophisticated and detailed equation of state, and we
employ the Boltzmann equation in our calculations of neutrino
heat transfer. Our results, especially as regards to the
survivability of fluctuations produced at very early epochs, can
differ from previous studies.

Fluctuations may be produced in the early universe with an
adiabatic, isothermal, or isocurvature character
(cf. Kolb \& Turner 1989).
It is conceivable that fluctuations could be produced which
have aspects of all of these characteristics. Adiabatic fluctuations
are not fluctuations in entropy-per-baryon, whereas, isothermal
or isocurvature fluctuations are entropy fluctuations. Once
fluctuations enter the horizon they evolve rapidly and nearly adiabatically
to pressure equilibrium with the background environment
(Hogan 1978). We will term any such fluctuation which has
come to pressure equilibrium as an {\it isobaric} fluctuation.

In Figure 1 we present a schematic picture for entropy fluctuation
classification and evolution. In this figure we show the ratio
of entropy density in a fluctuation to average entropy density
plotted against the ratio of baryon number density
in the fluctuation to average baryon number density.
All fluctuations evolve towards the isobaric line. Hydrodynamic
evolution tracks to pressure equilibrium for isothermal and
isocurvature fluctuations are shown as dotted lines.
Adiabatic fluctuations cease to exist when pressure equilibrium
is attained. Once pressure equilibrium for non-adiabatic
fluctuations is established subsequent evolution proceeds largely
along the isobaric line and is induced by neutrino, photon, and
baryon dissipative processes.

In Section (2.) we consider the physics of neutrino, photon, and
baryon dissipative processes. Section (3.) presents numerical
simulations for entropy fluctuation evolution. Conclusions
are given in Section (4.).

\vskip 0.15in
{\bf\centerline {2. Damping of Isobaric Entropy Fluctuations}}
\vskip 0.1in

In this section we discuss the evolution of subhorizon scale entropy
fluctuations. A schematic representation of the kind of fluctuation
we consider is shown in Figure 2. In this figure
$L^s$ is the mean separation between the centers of fluctuations
and $L$ is the width of a square wave fluctuation.
When fluctuations enter the horizon they undergo rapid hydrodynamic
expansion (or contraction) until they come into pressure equilibrium
with their surroundings. Fluctuations which are initially purely
adiabatic are completely erased by this process. Fluctuations which
have an initially isocurvature or isothermal component are not
erased by the expansion or contraction to pressure equilibrium.
We will call fluctuations which have attained pressure
equilibrium with their surroundings isobaric fluctuations.
In this paper we will consider the subsequent damping of isobaric
fluctuations by neutrino, baryon, and photon dissipative processes.

We define the baryon number overdensity distribution, $\Delta(x)$,
by
$$n_b(x)={\bar n_b}\bigl(1+\Delta (x)\bigr),\eqno(1a)$$
where $n_b(x)$ is the proper net baryon number density at space
coordinate $x$, and
$\bar n_b$ is the average proper net baryon number density
(see Figure 2).
We take the average proper number density at an epoch with scale
factor $R$ to be,
$${\bar n_b}(x)={\bar N_b}(x)\Bigl({R_{100}\over R}\Bigr)^3\
,\eqno(1b)$$
where ${\bar N_b}(x)$ is the average
proper number density at $T=100$ MeV,
and we take $R=R_{100}$ when $T=100$ MeV.
In what follows we choose $R_{100}=1$.

The distribution of energy density in relativistic particles,
$\varepsilon_{rel} (x)$, is defined in terms of the horizon average
of this quantity, ${\bar\varepsilon_{rel}}$ , by
$$\varepsilon_{rel}(x)={\bar\varepsilon_{rel}}\bigl(1+4
\delta (x)\bigr)\ .\eqno(2)$$
In this expression $4\delta (x)$ is then the deviation of the energy
density from its average value at position $x$.
We will be concerned with times early enough that
the cosmic energy density is dominated by highly relativistic particles.
The average energy density in relativistic particles in the dilute
high entropy conditions of the early universe is
$${\bar \varepsilon_{rel}}=g_{eff}({\bar T})a{\bar T^4}\ ,\eqno(3)$$
where $a=\pi^2/30$ , $\bar T$ is
the cosmic average temperature, and
$g_{eff}=\sum_bg_b+{7\over 8}\sum_fg_f$
are the number of
degrees of freedom in relativistic bosons ($b$) and fermions ($f$).
In this sum we will count only
those particles with mean free path, $l$,
smaller than the spatial dimension of the fluctuation.
This typical spatial dimension, or fluctuation length, we denote as
$L$ (see Figure 2).

In the limit where $\delta(x)<<1$ and in epochs where $g_{eff}$
can be approximated as constant in time, we find that
$$\delta (x)\approx{{T(x)-{\bar T}}\over {\bar T}}\ ,\eqno(4)$$
so that in this limit $\delta (x)$
is just the deviation from the cosmic
average temperature at coordinate $x$.

We can get a rough idea of how fluctuations come into pressure
equilibrium with their surroundings if we
assume that baryons contribute perfect
gas pressure
$P_b=n_bT$ and relativistic particles contribute pressure
$P_{rel}={1\over 3}\varepsilon_{rel}$.
In the following sections we will check the validity of these
assumptions and we will find significant
deviation from perfect gas pressure
at high temperatures $T$ $^>_{\sim} 30$ MeV.
Nevertheless, with these assumptions the demand for pressure
equilibrium
between the interior and exterior of a fluctuation implies,
$${1\over 3}{\bar\varepsilon_{rel}}\bigl(1+4\delta (x)\bigr)+
{\bar n_b}\bigl(1+\Delta (x)\bigr){\bar T}\bigl(
1+\delta (x)\bigr)\approx{1\over 3}{\bar\varepsilon_{rel}}+{{\bar
n_b}\bar T}\ .\eqno(5)$$
In the limit where $\delta << 1$ this expression reduces to
$$\delta (x)\approx-{1\over 4}\Biggl({{{\bar n}_b{\bar T}}
\over {1\over 3}{\bar\varepsilon_{rel}}}\Biggr)\Delta (x)\approx
-{1\over 4}
\Biggl({{\bar P_b}\over {\bar P_{rel}}}\Biggr)\Delta (x)\approx
-{\Delta (x)\over\bar s}\ ,\eqno(6)$$
with $\bar s$ the average entropy-per-baryon in the early universe.
Hereafter, a bar over any quantity denotes the average value of that
quantity over the whole horizon volume unless indicated otherwise.

In the early universe the average entropy-per-baryon is a
large quantity,
$$\bar s\approx 2.6\times 10^8\Omega_b^{-1}h^{-2}\ .\eqno(7)$$
Here $\Omega_b$ is the fraction of the closure density contributed by
baryons, and $h$ is the Hubble constant in units of
$100$ km s$^{-1}$ Mpc$^{-1}$.
The average entropy-per-baryon $\bar s$ is constant with time in the
limit where we can neglect baryon number violating processes
($T$ $^<_{\sim} 100$ GeV), black hole production and evaporation,
and dissipative processes. Note that $\bar s$ is large enough
that fluctuations will have $\delta <1$ for $\Delta$
$^<_{\sim}10^9$.

Pressure equilibrium for fluctuations
is obtained initially by hydrodynamic expansion.
The expansion of an entropy fluctuation can be rapid enough that
we can regard it as a nearly adiabatic process. After pressure
equilibrium is established we expect the temperature in the
fluctuation to be lower than $\bar T$ (see Figure 2).
Subsequent heat and entropy flow into the fluctuation
will tend to cause the fluctuation to expand
to a new pressure equilibrium. Entropy will flow into the fluctuation
until the average entropy-per-baryon in the fluctuation matches
the background average. Heat flow, or other dissipative processes,
will of course generate entropy. In practice, however, the entropy
generation due to dissipative processes is negligible for
fluctuations with initial $\delta <<1$.
The fluctuation will evolve through a
succession of pressure equilibrium states if the heat
transport time is long compared to a hydrodynamic expansion time.

The approximate time scale to attain pressure equilibrium between
the fluctuation and the environment
is the sound travel time, $\tau_p$,
over the fluctuation length $L$.
The time scale to transform an initially isobaric fluctuation into an
isothermal fluctuation is a typical heat transport time
scale $\tau_h$.
The ratio of the hydrodynamic expansion time scale
(or sound crossing time) to the typical heat transport time is,
$${\tau_p\over\tau_h}\sim{{L/v_s}\over {L^2/D_h}}\sim
{g_t\over g_{eff}}{l\over L}\ ,\eqno(8a)$$
where $v_s\approx c/\sqrt 3$ is the speed of sound for a relativistic
fluid. The heat diffusion constant is roughly
$D_h\sim c(g_t/g_{eff})l$, where $l$ is a typical
mean free path for heat-transporting particles, and
$g_t$ is the number of degrees of freedom which are
effective in heat transport. Equation (8a) is only valid in the
diffusive limit for heat transport, where $l<L$.

Particles with $l>L$ are free
streaming on the scale of the fluctuation. Even in this limit
particles can still transport heat into the fluctuation.
For neutrinos with $l>L$
we find a heating rate $\tau_h^{-1}\sim
(g_t/g_{eff})c/l$, which yields,
$${\tau_p\over\tau_h}\sim {g_t\over g_{eff}}{L\over l}
\ .\eqno(8b)$$

We conclude that the assumption that fluctuations evolve through a
succession
of pressure equilibrium configurations
is valid in the extreme
diffusive ($l<<L$) and \lq\lq homogeneous\rq\rq\ ($l>>L$)
heating limits.
Borrowing terminology from radiation transport studies
we can describe these limits as \lq\lq optically thick\rq\rq\
and \lq\lq optically thin\rq\rq , respectively.

For an intermediate mean free path, $l\sim L$, the heating and
pressure
equilibrium time scales become comparable, and the
calculation of the damping of
fluctuations by heat flow
would require a detailed solution of the Boltzmann equation.
However, in the case of neutrino heat transport in the early
universe we expect the approximation of pressure equilibrium to
hold fairly well, since neutrinos go quickly from optically
thick to optically thin on the scale of the fluctuation.
Furthermore the ratio $g_t/g_{eff}$ is a small quantity
for neutrino heat transport in the early universe,
which tends to make the ratio in equations (8ab) small.

Self gravity of subhorizon scale baryon number
fluctuations is negligible
compared to pressure forces in the early universe.
This is not surprising as the Jeans mass is close to the horizon
mass in a radiation dominated regime.
A rough estimate for the ratio of
pressure forces $F_p$ to gravitational forces $F_g$ on the edge of
a square wave fluctuation gives
$${F_p\over F_g}\sim 10^{16}\Delta^{-1}(\Omega h^2)^{-1}\biggl(
{L_{100}\over m}\biggr)^{-2}\ ,\eqno(9)$$
where $L_{100}$ is the fluctuation length $L$ measured on a
\lq\lq comoving \rq\rq\ scale at an epoch of $T=100$ MeV in a manner
to be described below.

We will frequently refer proper lengths to a
\lq\lq comoving\rq\rq\ scale at
$T=100$ MeV. The fluctuation length, $L$, and
separation length between
adjacent fluctuations, $L^s$, as measured on the $T=100$ MeV scale
are denoted by $L_{100}$
and $L^s_{100}$ respectively.
At an epoch where the temperature is $T$ and the scale factor
is $R(T)$ the relation between a proper length $L$ and the
associated \lq\lq comoving\rq\rq\ length $L_{100}$ is
$$L=L_{100}\biggl({R(T)\over R_{100}}\biggr)\ .\eqno(10)$$
If the product $RT$ were constant with time then we could
write $R(T)/R_{100}=$ $(100 MeV)/T$.
However, particle annihilation epochs change the value of
$RT$ (e.g. $e^{\pm}$-annihilation; $\mu^{\pm}$-annihilation;
quark-annihilation or QCD-transition). Figure 3 shows $R(T)(T/100
{\rm MeV})$
for $10^{-3}$ MeV $<T<$ $100$ MeV,
including a numerical treatment of all relevant annihilation
epochs. In this figure we assume that the quark-hadron transition
occurs at higher temperatures and we ignore pion degrees of freedom.
\vskip 0.15in

{\bf\leftline {a.) Neutrino Inflation}}
\vskip 0.1in

The neutrino mean free path $l_{\nu}$ is
a rapidly varying function of
temperature in the early universe,
$$l_{\nu}=\beta G_F^{-2} T^{-5}\approx 8\times 10^{-6}m
\biggl({T\over GeV}\biggr)^{-5}(N_l+3N_q)^{-1},\eqno(11)$$
where $G_F$ is the Fermi constant, and $N_l$ is the number of
relativistic weakly interacting leptons at temperature $T$.
These will include $e^{\pm}$ , $\mu^{\pm}$ , $\tau^{\pm}$ ,
$\nu_e$ , $\bar\nu_e$ , $\nu_{\mu}$ , $\bar\nu_{\mu}$ ,
$\nu_{\tau}$ , $\bar\nu_{\tau}$ , whenever their masses
satisfy $m_l<T$.
Similiary $N_q$ is the number of relativistic quark flavors
at temperature $T$ and may include $u$ , $d$ , $s$ , $c$ , $b$ ,
and $t$.
For example, at a temperature below that for quark annhilation
$N_l\approx 10$ ( $e^{\pm}$, $\mu^{\pm}$, and six neutrino species )
and $N_q=0$. Here we have neglected neutrino-pion scattering.
The dimensionless
quantity $\beta$
depends on neutrino flavor, the types of relativistic leptons
(quarks) in equilibrium at temperature
$T$, and the exact lepton (quark) weak couplings.
Heckler and Hogan (1993) have considered neutrino scattering
in a weakly coupled relativistic plasma at zero chemical
potential. They found that plasma screening corrections
are negligible, so that neutrinos can be approximated
as scattering incoherently off single particles.

Neutrinos are very efficient heat transporters at any time
in the early universe between
the epochs of the electroweak
phase transition ($T\approx 100$ GeV)
and neutrino decoupling ($T\approx 1$ MeV).
Before the electroweak phase transition neutrino heat transport is
inefficient, since for vanishing weak gauge boson masses
($M_{Z,W}=0$) the neutrino mean free path becomes small.
Since the neutrino mean free path changes considerably
over the history of the universe where we follow entropy
fluctuations we
will
have to treat neutrino heat
transport in two different limits: the homogeneous limit,
$l_{\nu}>L$;
and the diffusive limit, $l_{\nu}<L$.

In the homogeneous limit neutrinos are free streaming through
the fluctuation, so that neutrinos are optically thin on the scale
of the fluctuation $L$. This implies that the neutrino energy
density is homogeneous on these scales with
$\varepsilon_{\nu}(x)=\bar\varepsilon_{\nu}$. In other words,
the temperature for neutrinos is now slightly larger than the temperature for
other relativistic particles inside the fluctuation.
In this case neutrino scattering will tend to transfer energy into the
fluctuation.
The heat deposition rate in a fluctuation of size $L$ is equal to the
number of scattering events per unit time multiplied by the
average energy transfer per collision.
The scattering event rate is the
neutrino flux ($\sim (g_{\nu}/10)T^3$) times the weak cross section
($\sim G_F^2T^2$) times the total number of targets within the
fluctuation ($\sim (7/8)(2/10)(N_l+3N_q)T^3L^3$),
assuming natural units $c=\hbar =1$.

The fractional difference in temperature between the neutrinos
and the rest of the particles in the fluctuation at position $x$
is just $\delta (x)$. The average energy transfer per collision at
position $x$
is then roughly $\delta (x)T_{\nu}$, where $T_{\nu}$ is the
homogeneous neutrino distribution temperature. Note that
$T_{\nu}=\bar T$. These arguments yield the rate of
change of energy density in the fluctuation,
$${\partial\varepsilon (x)\over\partial t}\approx -FG_F^2{\delta (x)}
{\bar T}^9\ ,\eqno(12)$$
where $F$ is a numerical factor.
At temperatures $T$ $^<_{\sim}$ 50 MeV we can approximate
$F\approx (g_eg_{\nu}/100)\approx 0.2$, where $g_e=(7/8)2N_e$,
$N_e=2$, and
$g_{\nu}=(7/8)6$. This naive estimate for the homogeneous neutrino
heating rate turns out to be low compared to that derived from our
detailed numerical treatment.

We have computed the homogeneous neutrino heating
rate by using the Boltzmann transport equation.
Specifically, we evaluated the collision integral term in the
Boltzmann equation.
In this calculation we solved for the energy transfer between a
relativistic neutrino Fermi-Dirac distibution at temperature $\bar T$
and a relativistic Fermi-Dirac $e^{\pm}$ distribution at a lower temperature
${\bar T}(1-\delta )$ .
The calculation neglects Fermi-blocking effects on scattering
particles which is a somewhat reasonable approximation in the early universe.
Our numerical result has the same form as equation (12) but has
$F=1.869$, so that the net heating rate is nearly an order of
magnitude larger than the simple estimate. This result is not
surprising, however, since the naive estimate of equation (12) does not
take account of the contribution to heating from annihilation
processes such as $e^+e^-\leftrightarrow \nu {\bar\nu}$.
We include these processes in our numerical calculation.
These processes dominate the energy deposition, accounting for
$70\%$ of $F$.

In the limit where $l_{\nu}>L$ we can employ equations (2) and (3)
to rewrite equation (12) as
$${d\delta (x)\over dt}\approx -F^{\star}{g_{\nu}\over g_{eff}}
{1\over
l_{\nu}}\delta (x)\ ,\eqno(13a)$$
where,
$$F^{\star}={\beta\over g_{\nu}}{1\over 4a}F\ .\eqno(13b)$$
Note that $F^{\star}$ is roughly independent of $g_{\nu}$, $N_l$,
and $N_q$.

When $l_{\nu}<L$ the optically thick limit obtains and we can
approximate the neutrino heating as arising from neutrino diffusion.
In this limit we can express the time evolution of $\delta (x)$
for a spherical fluctuation as,
$${\partial\delta (r_{100})\over\partial t}
={D_h\over R^2}{1\over r_{100}^2}
{\partial\over\partial r_{100}}\Biggl(r_{100}^2
{\partial\delta (r_{100})\over\partial r_{100}}\Biggr)\ ,\eqno(14)$$
where $r_{100}$ is the radial coordinate of a spherical fluctuation
measured on our comoving scale at $T=100$ MeV, and
$D_h\sim (g_{\nu}/g_{eff})l_{\nu}$ is the heat diffusion constant.
For a two-zone square wave
fluctuation in the limit where $l_{\nu}<L$ we find
$${d\delta\over dt}=-\lambda^{\star}{g_{\nu}\over g_{eff}}
{l_{\nu}\over L^2}\delta\ ,\eqno(15)$$
with $\lambda^{\star}$ a numerical factor of order unity.
In this two-zone approximation for the density and temperature distribution
of a spherical fluctuation $4\delta$ represents the fractional energy
density difference between the inside (zone 1) and the outside (zone 2)
of the fluctuation.
Matching the heating rates
in the diffusive and homogeneous limits for $l_{\nu}=L$ we
obtain $\lambda^{\star}=F^{\star}$.

We now consider a two-zone square wave fluctuation, which will serve
to illustrate the typical damping histories for fluctuations.
A simple prescription for calculating the damping
of isobaric square wave
baryon number fluctuations by neutrino heat transport is as follows.
If in time
$t$ a fluctuation is heated by $d\delta$, it has to
adiabatically expand by $dL_{100}=L_{100}d\delta$
in order to re-establish
pressure equilibrium. Thus,
$${1\over L_{100}}{dL_{100}\over dt}={d\delta\over dt}\ ,\eqno(16)$$
where $L_{100}=L/R$ as before.
Baryon number within the fluctuation is conserved during the
adiabatic expansion of the fluctuation
and so the
quantity $(1+\Delta)L_{100}^3$ is time independent.
In order to obtain the rate of increase in radius
(or \lq\lq inflation\rq\rq\ rate) of isobaric square wave
baryon number fluctuations
we integrate equation (16), using
equation (13) in the homogeneous limit, and equation (15)
in the diffusive limit. To accomplish this we need to know the
temperature difference, $\delta$, as a function of the baryon
number overdensity, $\Delta$, for a fluctuation in pressure
equilibrium.

The pressure due to relativistic particles can be written
$$P_{rel}=f_p\varepsilon_{rel}\ ,\eqno(17)$$
with $f_p=1/3$, except during epochs of lepton- or quark-
annihilations.
Figure 4 shows $f_p(T)$ during the $e^+e^-$ annhilation. $f_p$ drops
from $f_p=1/3$ to $f_p=0.28$ at $T\approx 0.2$ MeV. At
this temperature a major fraction of kinetic energy of the
relativistic gas is converted to rest mass energy of
$e^+e^-$ pairs.
This rest mass makes zero contribution to the pressure.

We consider four temperature regimes in which there are natural
limits for the pressure contributed by baryons.
At temperatures below the QCD-transition baryon number is carried
by color singlet baryonic states. In the regime after the QCD-transition
the kinematics of baryons can be taken to be approximately nonrelativistic.
Below $T\approx 30$ MeV baryons exert
perfect gas pressure $P_b$
$$P_b=n_bT\approx {\bar n}_b{\bar T}(1+\Delta)\ .\eqno(18)$$
The pressure contribution of
the excess electrons needed for charge neutrality
is negligible compared to the baryon pressure.
The entropy-per-baryon is so high that
the net number of electrons is small
compared to the number of electron-positron pairs.
Requiring pressure equilibrium as we did in equation (6), we obtain
for $T$ $^<_{\sim}$ 30 MeV,
$$\delta\approx -{1\over 4}\Biggl({{\bar n}_bT\over f_p
{\bar\varepsilon}_{rel}}\Biggr)\Delta\ .\eqno(19)$$

In Figure 4 we show the proportionality constant between
$\Delta$ and $\delta$ through the $e^+e^-$ annihilation epoch.
In order to keep the fluctuation in pressure
equilibrium during the $e^+e^-$ annihilation process $\delta$
has to
increase. An increase in $\delta$ enhances heat flow into
the fluctuation. During the $e^+e^-$ annihilation at
$T\sim 0.5-0.05$ MeV there are no particles which are efficient
heat transporters, so that the pair-annihilation effect is
unimportant.
However, during epochs of quark annihilation at high temperatures
or muon and pion annhilation at $T\sim 100$ MeV,
the neutrino is an efficient heat transporter and heat
transport efficiency is slightly enhanced during the annihilation epochs.

Thermally produced baryon-antibaryon pairs are still abundant enough
for $T$ $^>_{\sim}$ 30 MeV
to modify the pressure from the perfect gas limit in equation (18).
We find
$$P_b\approx T\sqrt{n_{pair}^2+n_b^2}\ ,\eqno(20)$$
where
$$n_{pair}\approx\biggl({2\over\pi}\biggr)
^{3\over 2}(mT)^{3\over 2}exp\Bigl(
-{m\over T}\Bigr)\Bigl(1+{15\over 8}{T\over m}\Bigr)\eqno(21)$$
is the total proton (neutron) plus antiproton (antineutron) particle
density of a zero chemical potential Fermi-gas at temperature $T$.
The baryon mass is taken to be $m$.
The overpressure exerted by the net baryon number
falls below perfect gas
pressure, since increasing the baryon number of a $\mu_b=0$
Fermi gas by two is equivalent to adding one baryon and
annihilating one antibaryon.
For a temperature approaching the
Hagedorn temperature $T_H\sim 140$ MeV (Hagedorn 1971),
the actual baryonic particle density exceeds the expression for
$n_{pair}$ in equation (21) due to
the existence of thermally produced baryonic resonances.
Since baryon number can be carried by additional degrees of freedom,
the resulting overpressure due to extra net baryon number
is further decreased for $T\mapsto T_H$ (Alcock, Fuller, \& Mathews 1987).
Note that equation (20)
reduces to the perfect gas pressure limit of equation (18), when
$n_{pair}<<n_b$.

We can estimate a temperature difference $\delta$ between
the inside of a fluctuation and the
cosmic background environment from pressure equilibrium,
$$\delta=-{1\over 4}{{\bar T}\over f_p{\bar\varepsilon}_{rel}}
\biggl(\sqrt{n_{pair}^2+{\bar n}_b^2\bigl(1+\Delta
\bigr)^2}
-\sqrt{n_{pair}^2+{\bar n}_b^2}\biggr)\ .
\eqno(22)$$
In Figure 5 we illustrate the decrease of fluctuation temperature
difference $\delta$ at high temperatures for three different
fluctuation overdensities $\Delta$.
The results in this figure do not include baryonic resonances.
We show the ratio
$\delta/\delta_{pg}$, with $\delta_{pg}$ the temperature difference
derived using
perfect gas pressure for baryons but with no
baryon-antibaryon pairs.
It is interesting to note that at high temperatures, but after the
QCD-transition, fluctuations with net baryon number ($\mu_b\neq 0$)
can {\it almost} coexist in thermodynamic
equilibrium with a vanishing chemical
potential ($\mu_b=0$) phase.

For times earlier than the QCD-transition baryon number is carried
by relativistic quarks.
The QCD-equation of state in the
strong QCD-coupling limit ($\alpha_s\sim 1$) is uncertain.
The strong coupling limit obtains for $T$ $^<_{\sim}$ 3-5 GeV
(Mueller 1985).
In the weak QCD-coupling limit
($\alpha_s<<1$) we can approximate
the equation of state as that of a relativistic
ideal Fermi gas with net baryon number density $n_b$
and small baryon chemical potential $\mu_b<<T$,
$$P_q={7\over 4}N_qaT^4+{9}N_q^{-1}{n_b^2\over T^2}\ ,
\eqno(23)$$
where $N_q$ are the number of relativistic quark flavors, and $P_q$
is the total pressure of quark-antiquark pairs
plus net baryon number.
The overpressure due to net baryon number is the second
term in equation (23).
First order perturbative strong interaction
corrections would increase this term by a
factor $(1-2\alpha_s/\pi)^{-1}$ (Mueller 1985).
Using equation (23) we obtain for the temperature difference
between the interior and exterior of a two-zone
fluctuation
$$\delta=-{9\over 4f_p{\bar\varepsilon}_{rel}}
{{\bar n}_b^2\over {\bar T}^2}
{1\over N_q}\bigl((\Delta+1)^2-1\bigr)\
 .\eqno(24)$$

We are now in the position
to integrate equation (16) and find by how much a fluctuation
inflates.
We take $F^{\star}=0.747$ and
$\beta =2.762$ and we assume that these quantities are approximately
temperature independent.
The QCD-transition is taken to be at $T=100$ MeV and the
electroweak phase transition is taken to occur at a temperature
above $T=100$ GeV. Our estimate
treats all quark- and lepton- annihilations as occuring
instantaneously
at temperature $T=m$.
Quark masses are taken here to be $m_u=m_d=0$, $m_s=100$ MeV,
$m_c=1.35$ GeV, $m_b= 5$ GeV and
$m_t> 100$ GeV.
We take the muon mass as $m_{\mu}=105.7$ and the $\tau$-lepton
mass as $m_{\tau}=1.78$ GeV.
We take account of the increase of scale factor
times temperature ($RT$) and decrease of statistical
weight, $g$, during the
annihilation epochs. The scale factor $R$ depends on temperature as
$$R=\biggl({g_s\over 10.75}\biggr)^{-{1\over 3}}\biggl({T\over 100 MeV}
\biggr)^{-1},\eqno(25)$$
assuming complete pion and muon annihilation at
$T=100$ MeV ($g_s=10.75$).
The time-temperature relationship is
$$t=2.42s\quad g^{-{1\over 2}}\biggl({T\over
MeV}\biggr)^{-2},\eqno(26)$$
for a simple radiation dominated expansion.
For the early universe we calculate an average net baryon number
density,
$${\bar n}_b=1.7\times 10^{-9}g_s\ \Omega_b\ h^2{\bar T}^3.\eqno(27)$$
In equation (26) we define the total statistical weight in
relativistic particles to be
$g=\sum_bg_b(T_b/T)^4+(7/8)\sum_fg_f(T_f/T)^4$, where we recognize
that decoupled particles may have temperatures ($T_b,T_f$) which
differ from the plasma temperature ($T$). The relevant statistical
weight for entropy density, $g_s$, enters into equations
(25) and (27) and is defined as
$g_s=\sum_bg_b(T_b/T)^3+(7/8)\sum_fg_f(T_f/T)^3$.

We have done numerical calculations for two-zone square wave
fluctuations based on the above approximations. We have also
generalized this calculation to include many zones of different
$n_b(x)$ and thus obtained a better approximation to the
continuum limit.
We will first discuss our results for two zones and then will
present our results for the multi-zone calculation in Section 3.

In Figure 6 we show the evolution of
neutrino mean free path $l_{\nu}^{100}=l_{\nu}/R$
and fluctuation length
$L_{100}=L/R$ for three different overdensities, assuming an
initial fluctuation length $L_{100}^i=10^{-8}m$ .
The proper fluctuation length
of very high amplitude fluctuations increases
at almost the same rate as the neutrino mean free path
until the overdensity $\Delta$ is reduced by several
orders of magnitude and the neutrinos become optically thin on
the scale of the fluctuation.
Therefore, most of the damping of nonlinear entropy fluctuations occurs
in the diffusive limit ($l_{\nu}<L$).
Note that the neutrino mean free path increase,
$dl_{\nu}/dt$, approaches the speed of light near neutrino decoupling
($T\approx 1$ MeV).
The neutrino mean free path and the fluctuation length appear
slightly discontinous at $T=5$ GeV because of our oversimplified
treatment of $b\bar b$-annihilation. However, the discontinuities
are small because of the large number of degrees of freedom at this
epoch in the early universe.
The results of an exact treatment of annihilation for a single
degree of freedom would not differ appreciably from our
estimates.

Figure 7 shows the results of our two-zone calculation for
the damping of fluctuations by neutrino inflation
between $T=100$ GeV and $T= 100$ MeV. We display the final
baryon-to-entropy ratio $(n_b/s)_f$ of inflated fluctuations as
a function of initial fluctuation length $L_{100}^i$
for five different initial
baryon-to-entropy ratios $(n_b/s)_i$.
We find considerable convergence in $(n_b/s)_f$
for a broad range of initial amplitudes, $(n_b/s)_i$, when
fluctuations have
small initial length scales. Fluctuations with very small
length scales $L_{100}^i$ $^<_{\sim}$ $10^{-14}$ m will be damped to
a characteristic $(n_b/s)_f\approx 2\times 10^{-8}$.
Fluctuations with large initial length scales show little damping
from neutrino inflation. Neutrino inflation becomes an
important damping process for fluctuations with initial length
scales smaller than some critical length. This critical length
decreases as the initial fluctuation amplitude decreases.
In Figure 7 the electroweak epoch horizon scale ($T=100$ GeV)
is approximately at the far right hand end of the horizontal
axis.
Large overdensity fluctuations with large initial length scales
have significant damping from neutrinos
during the strong QCD-coupling epoch.
The uncertainties in the equation of state at this epoch translate
into uncertainties in our damping estimates. We show our results
as dashed lines in this uncertain damping regime.

In Figure 8 we display our results for neutrino inflation
of fluctuations between
$T=100$ MeV and $T=1$ MeV. In this calculation we have neglected
the effects of baryonic and mesonic resonances.
Fluctuations with small initial length scales
($L_{100}^i$ $^<_{\sim}$ 0.1 m)
converge to a
final baryon-to-entropy ratio of $(n_b/s)_f\approx 1.1\times
10^{-5}$.
This corresponds to a baryon-to-photon ratio inside the fluctuation
of $\eta\approx 8\times 10^{-5}$ during primordial nucleosynthesis.
Fluctuations with initial length scales larger than
($L_{100}^i$ $^>_{\sim}$ 1 m) have less damping
since neutrinos become optically thin on the scale of the fluctuation
at low temperatures.
The dotted line divides fluctuations containing less (left)
and more (right)
baryonic mass than the total baryon mass in the horizon at the
QCD-transition epoch ($T=100$ MeV).

The effects of neutrino damping on fluctuations between
temperatures of $T=100$ GeV
and $T=1$ MeV are shown in Figure 9.
This figure combines results from Figure 7 and 8.
Our calculation assumes no fluctuation modification
from phase transition effects
during the
QCD-transition epoch.
The dashed lines indicates uncertainties in the
damping
due to equation of state uncertainties associated with
the strong QCD-coupling regime.
Even if we were to assume no damping of entropy fluctuations
in the strong QCD-coupling regime
Figure 9 would be essentially unchanged.
Significant damping of entropy fluctuations in the strong
QCD-coupling limit is unlikely because of the small neutrino mean
free path during this epoch.
It is evident that neutrino
inflation results in
a high degree of convergence in the final
baryon-to-entropy ratio
for a broad range of initial fluctuation amplitudes and length
scales. If neutrino inflation operated only in the
homogeneous limit
we would expect fluctuations of any initial length scale to
converge to the same $(n_b/s)_f$. Most of the neutrino
inflation for fluctuations with length scales in the range
$10^{-1}$m $ ^>_{\sim}\ L^i_{100}$ $ ^>_{\sim}10^{-5}$ m
takes place in the homogeneous regime so that we see
a fair degree of convergence in $(n_b/s)_f$ for these
fluctuations in Figures 8 and 9.

How sensitive are these results to our assumptions about the
microphysical
quantities $\beta$ and $F^{\star}$, the expansion rate of the
universe, and our simplified treatment of annihilation
epochs and phase transitions?
When there is sufficient inflation
($\Delta_f$ $^<_{\sim}  (1/10)\Delta_i$ with
$\Delta_f$, $\Delta_i$ the final- and initial- baryon number
overdensities) from high initial temperatures ($T$ $^>_{\sim}$ 1 GeV)
we can obtain an approximate analytic
result for the final baryon-to-entropy ratio
$$\Bigl({n_b\over s}\Bigr)_f\approx 4.5\times 10^{-7}
A
\Bigl({g\over 100}\Bigr)^{1\over 9}
\Bigl({g_{\nu}\over {7\over 8}6}\Bigr)^{-{4\over 9}}
\Bigl({N_q^4\over {N_l+3N_q}}\Bigr)^{1\over 9}
\Bigl({\beta\over 2.762}\Bigr)^{1\over 9}
\Bigl({F^{\star}\over 0.747}\Bigr)^{-{4\over 9}};\eqno(28a)$$
where,
$$A=\Bigl({\Delta_i\over 10^6}\Bigr)^{1\over 9}
\Bigl({L_{100}^i\over 10^{-10}m}\Bigr)^{1\over 3}
\Bigl({\Omega_b h^2\over 0.01}\Bigr)^{1\over 9}.\eqno(28b)$$
Quantities in equations (28ab), which are temperature dependent
(e.g. $g$, $N_q$, and $N_l$) should be
taken to be their values at the epoch where the neutrino mean
free path is of the order of the fluctuation length scale.
The $(n_b/s)_f$ in equation (28ab) is only very weakly
dependent on the
mean free path ($\sim\beta$), the total statistical weight ($g$),
and the number of relativistic quarks $N_q$ and leptons
$N_l$ at temperature $T$.
Furthermore, the effects of neutrino inflation are clearly
insensitive to
the average baryon-to-photon ratio
($\sim\Omega_b h^2$) and the initial
overdensity $\Delta_i$.
There is more significant dependence of the final ratio  $(n_b/s)_f$
on the number of heat
transporters ($g_{\nu}$) and the efficiency of the heat
transport ($\sim F^{\star}$).
Uncertainties in the precise QCD-transition temperature will not
affect the outcome of
neutrino inflation significantly unless the QCD-transition
temperature were very low ($T<100$ MeV).
Neutrino inflation is not very efficient
after the QCD-transition
if there are large numbers of thermally produced
baryon-antibaryon pairs (Figure 5)
as would be the case for a QCD-transition temperature $T>100$ MeV.
We do not
expect the annihilation of pions and muons to substantially
modify our results.

\vskip 0.15in
{\bf\leftline {b.) Photon Inflation and Hydrodynamic Expansion}}
\vskip 0.1in

At temperatures lower than $T\approx 30$ keV
baryon-to-entropy fluctuations
are efficiently damped by photon diffusion and hydrodynamic expansion against
photon-Thomson-drag (Alcock et.al. 1990).
The number density of $e^{\pm}$-pairs in the
universe remains large compared to the number density of ionization
electrons until the temperature drops below about
$T\approx 30$ keV.
The photon mean free path or, equivalently,
the total photon transport
cross section is determined by photon-electron and photon-positron
scattering. When the pairs finally annihilate we expect a substantial
increase in the photon mean free path. If the photon mean free
path becomes of order or larger than the fluctuation size, $L$,
then the photons go from optically thick to optically thin on this
scale. In the optically thin limit all temperature differences
between the inside and outside of the fluctuation are erased.
In this case we can no longer use radiation pressure
to balance the difference in baryon pressure
between the inside and outside of the fluctuation. The unbalanced
baryon pressure leads to hydrodynamic expansion of the fluctuation.

In the limit where Klein-Nishima corrections are small
($T$ $^<_{\sim}$ $m_e$)
the photon mean free path is
$$l_{\gamma e}\approx {1\over\sigma_Tn_{e^{\pm}}}\ ,\eqno(29)$$
where $\sigma_T=6.7\times 10^{-25}$ cm$^2$
is the Thomson cross section,
and $n_{e^{\pm}}=n_{e^-}+n_{e^+}$
is the number density of electrons plus positrons.
At temperatures above the electron rest mass the
photon-electron scattering cross section is smaller than the Thomson
cross section. However, the photon mean free path is still small
in this case since there are a
large number of thermally produced charged particles.
Figure 10 shows the photon mean free path
$l_{\gamma}^{100}=l_{\gamma}/R$
for temperatures below $T=1$ MeV.
Figure 3 for $R(T)$ can be used to convert comoving distances
into proper distances.

In Figure 10 we show the photon mean free path for three
different proton fluctuation amplitudes $\Delta_p$.
The proton density, $n_p$, which equals the net electron density
from charge neutrality, is just
$$n_p={\bar n_b}\Delta_p\ .\eqno(30)$$
In Figure 10 we neglect alpha particles.
The photon mean free path rises sharply during the epoch of
$e^+e^-$-annihilation until the number density of thermal
$e^+e^-$-pairs drops below the net electron density at
temperatures around $T\approx 30$ keV.
After $e^+e^-$-annihilation the photon mean free path depends only
on the net electron density and so tracks the proton density
across a fluctuation. Thus the photon mean free path can vary by
several orders of magnitude across a fluctuation.
A large photon mean free path contributes to efficient
damping of
small-scale nonlinear
entropy fluctuations. We consider two limits for damping of
fluctuations by this process which are the analogues of the two
limits considered for neutrino inflation.

Fluctuations expand due to diffusive photon heat transport
when the photon mean free path is smaller than the fluctuation
length scale $l_{\gamma}<L$. In this limit the heat diffusion
equation (14) applies, with the photon heat diffusion
constant $D_{\gamma}$ approximately given by
$$D_{\gamma}\approx {g_t\over g_{eff}}l_{\gamma e}\ ,\eqno(31)$$
where $g_t$ is the statistical weight of heat transporting
particles (photons, $g_t=2$) ,
and $g_{eff}$ is the statistical weight of relativistic particles
still
\lq\lq coupled\rq\rq\ to the material in the fluctuation
($\gamma,e^{\pm}$).
Using equations (14) and (16) we can obtain an approximate damping
time scale. We take this to be $\tau_{\gamma}$,
the time to double the size of a square wave
entropy fluctuation by photon heat advection,
$$\tau_{\gamma}^{-1}\approx
{1\over 5\times 10^3s}\biggl({L_{100}\over m}\biggr)^{-1}
\biggl({T\over
10 keV}\biggr)^{-1}\biggl({{2\Delta_p+3\Delta_{^4He}}\over
\Delta_p+2\Delta_{^4He}}\biggr)\ ,
\eqno(32)$$
where $L_{100}$ is the comoving fluctuation length as before,
and $\Delta_{^4He}$ is the $^4He$ density
defined in a manner analogous to
$\Delta_p$ in equation (30). Since we here deal with temperatures well
within the nucleosynthesis epoch we must allow for a significant
$^4He$ mass fraction.
Equation (32) assumes complete annihilation of $e^+e^-$-pairs
and thus applies only at low temperatures $T$ $^<_{\sim}$ 20 keV.
The denominator in the last factor of equation (32) arises from
the net electron density within the fluctuation, whereas the
numerator is related to the temperature deviation, $\delta$, in the
fluctuation.

When the photons become optically thin on the scale of the
fluctuation ($l_{\gamma}>L$) then the temperature across the
fluctuation becomes uniform
and the fluctuation disintegrates by
hydrodynamic expansion.
The overpressure due to extra baryons and electrons within the
fluctuation will drive collective fluid expansion out of the
fluctuation. However, protons moving through an
isotropic photon gas will experience a drag force proportional
to the fluid velocity. As seen from the proton rest frame
the photon flux against the direction of motion is slightly
higher than the photon flux from the opposite direction.
Electrons are dragged along by the protons through electric forces.
These electrons
Compton scatter
more often in head-on collisions transfering some net
momentum to the photon gas.
This leads to the Thomson drag force on the moving fluid
$F=(4/3)\sigma_T\varepsilon_{\gamma}
(v/c)$ (Peebles 1971). Because of this drag force the
expanding fluid will reach a terminal peculiar velocity
$v=(dr_{100}/dt)$,
$$v\approx {3\over 4\sigma_T\varepsilon_{\gamma}n_e}{1\over R^2}
{dP\over dr_{100}},\eqno(33)$$
where $dP/dr_{100}$ is the radial pressure gradient,
$r_{100}$ is the radial coordinate as measured on our comoving
scale,
$\varepsilon_{\gamma}$ is
the energy density in photons, and $n_e=n_{e^-}-n_{e^+}$
is the net electron density.
We find for the pressure exerted by baryons and electrons
below $T\approx 30$ keV
$$P\approx {\bar T}{\bar n_b}\Biggl(\sum_i\Delta_i+\biggl(
{n^{*}_{pair}}^2+\Bigl(\sum_iZ_i\Delta_i\Bigr)^2\biggr)^
{1\over 2}-n^{*}_{pair}\Biggr)\ ,\eqno(34)$$
with the sum running over all nuclei $i$ with nuclear charge
$Z_i$. In this expression the $e^+e^-$-pair density divided by
$\bar n_b$ is
$$n^{*}_{pair}\approx {1\over\bar n_b}{1\over 2}\biggl({2\over\pi}
\biggr)^{3\over 2}(m_eT)^{3\over 2}exp\Bigl(-{m_e\over T}\Bigr)
\Bigl(1+{15\over 8}{T\over m_e}\Bigr)\ ,\eqno(35)$$
which is similiar to equation (21) for
baryons.
The time scale to double the size of
a square wave fluctuation by hydrodynamic expansion
can be obtained from a characteristic fluid
velocity and a characteristic fluctuation length $L_{100}$.
This time scale is
roughly $\tau_h\approx L_{100}/v$.
After complete $e^+e^-$-annihilation
we find
$$\tau_h^{-1}\approx{1\over 2.4\times 10^3s}
\biggl({L_{100}\over m}\biggr)^{-2}
\biggl({T\over 10 keV}\biggr)^{-1}\biggl({{2\Delta_p+
3\Delta_{^4He}}\over\Delta_p+2\Delta_{^4He}}\biggr)\ ,\eqno(36)$$
with the notation as before. This hydrodynamic expansion time scale
is similiar to the inflation time scale from diffusive photon heat
transport in equation (32).
It is important to note that the damping time scales for diffusive
photon heat flow and hydrodynamic expansion are independent
of fluctuation amplitude.
In Figure 11 we show damping rates for diffusive photon inflation,
$\tau_{\gamma}^{-1}$ (dashed), and hydrodynamic expansion,
$\tau_h^{-1}$ (dashed-dotted), for a fluctuation with $L_{100}=1$ m.
In this figure we also show the reciprocal of the Hubble
time $\tau_H^{-1}\approx (8\pi G\varepsilon/3)^{1/2}$ (solid)
between temperatures
of $T=100$ keV and $T=1$ keV.
Damping time scales are shorter than the Hubble time at this epoch
for fluctuations smaller than $L_{100}$ $^<_{\sim}$ 1 m and for
temperatures below $T\approx 30$ keV. We conclude that small
inhomogeneities
will disintegrate rapidly below $T\approx 30$ keV.
Larger fluctuations, $L_{100}>1m$, will
disintegrate at somewhat lower temperatures.

In this work we ignore all nonspherical dissipative modes
(e.g. convection, dendritic instabilities, and
Rayleigh-Taylor instabilities).
This approximation is valid over all the range of conditions
we consider except possibly at very late times when rapid
hydrodynamic expansion against Thomson drag occurs. Simple
estimates show that the acceleration at these times is so small
that the surface of an expanding fluctuation is Raleigh-Taylor
stable.

\vskip 0.15in
{\bf\leftline {c.) Baryon Diffusion}}
\vskip 0.1in

High-amplitude subhorizon-scale baryon-to-entropy fluctuations
in the early universe are modified considerably by baryons
diffusing through the primordial plasma
(Applegate, Hogan, and Scherrer 1987).
The baryon diffusion equation in spherical comoving
coordinates, $r_{100}=r/R$, is
$${\partial\Delta_i\over\partial t}={1\over r_{100}^2}
{\partial\over\partial r_{100}}\biggl({D_i\over R^2}r_{100}^2
{\partial\over\partial r_{100}}\Delta_i\biggr)-\sum_j
f_{ij}(T)\Delta_j\ ,\eqno(37)$$
where $\Delta_i$ is the ratio of the number density of nucleus $i$
to the
total baryon number density,
and $D_i$ is the diffusion constant for the ion corresponding to
nucleus $i$. We include in equation (37)
a source term representing changes in the nuclear abundances
due to nuclear reactions, so that the coefficients $f_{ij}$ represent
nuclear reaction rates.
Up to a numerical factor of order
unity the baryon diffusion constant $D_i$ for nucleus $i$ is
$$D_i\approx {1\over 3}v_il_i=
{1\over 3}v_i{1\over\sigma_{ik}n_{ik}}\ ,\eqno(38)$$
with $v_i$ the thermal baryon velocity, $l_i$ the baryon
mean free path,
$\sigma_{ij}$ the cross section for the scattering
of nucleus $i$ on particle $k$,
and $n_k$ the number density of particle $k$.

In Table 1 we list the baryon diffusion constants for neutrons and
protons, as well as the heat diffusion constant for photons in
the early universe
(Applegate, Hogan, \& Scherrer 1987; Banerjee \& Chitre 1991;
Gould 1993).
We show baryon diffusion constants for the
scattering of baryons on individual particle species $k$.
The effective baryon diffusion constant of nucleus $i$ in the
primordial plasma due to all scattering is
$${1\over D_i}=\sum_j{1\over D_{ij}}\ .\eqno(39)$$
Table 1 also displays the relevant cross sections and the
relative cosmological importance of individual scattering
processes.

Neutrons diffuse most
easily through the primordial plasma.
At temperatures above $T\approx 100$ keV
neutron diffusion is
limited by magnetic moment scattering on electrons and positrons.
However, at lower
temperatures ($T$ $^<_{\sim}$ 100 keV),
or in high-density fluctuations
with a local baryon-to-photon ratio $\eta$
$^>_{\sim}$ $3\times 10^{-8}$,
neutron diffusion is limited by nuclear scattering on protons.
Thus, the diffusive damping of high-amplitude fluctuations
($\Delta>>1$)
by baryon diffusion is a nonlinear process,
since for a higher proton density neutron diffusion is less
efficient.

Before the freeze-out from weak equilibrium ($T\approx 1$ MeV)
there is no segregation between
neutrons and protons, and baryons
only diffuse efficiently during the fraction of time they spend as neutrons.
After weak equilibrium freeze-out protons and neutrons diffuse
independently, and since protons and neutrons have different diffusion
constants the result will be spatial segregation of these species.
Proton diffusion has negligible effects above $T\approx 30$ keV, except for
the very smallest fluctuations $L_{100}<< 1m$. However, such small
fluctuations are damped efficiently by baryon diffusion
before weak freeze-out.

For temperatures above $T\approx 40$ keV Coulomb scattering of protons
off electrons and positrons dominates the proton diffusion constant.
This is because protons move through the plasma as independent
particles, in particular protons do not drag along electrons. The
proton electric charge is shielded by plasma electrons since the
Debeye screening length is much smaller than a typical interparticle
spacing (Applegate, Hogan, and Scherrer 1987).
The diffusivity of heavy nuclei for temperatures above $T\approx 40$
keV should as well be limited by Coulomb scattering of the nuclei
off electrons and positrons. A simple estimate for the diffusion
constant of nuclei relative to that of protons can be easily
obtained from equation (38). The diffusivity of nuclei should be
suppressed relative to the diffusivity of protons because of a
larger Coulomb cross section for the scattering of $e^+e^-$ off
nuclei and a smaller thermal velocity of the nuclei. For a nucleus
with nuclear charge $Z_i$ and nucleon number $A_i$ the Coulomb cross
section is $Z_i^2$ times larger than that for protons and the
thermal velocity is $1/A_i$ times smaller than the proton thermal
velocity. This leads to a suppression factor of the diffusivity of
nuclei relative to that for protons of $(1/Z_i^2A_i)$.

In the temperature range $40$ keV $^>_{\sim}$ $T$
$ ^>_{\sim}$ 0.1 keV Compton
scattering of electrons off photons limits proton diffusion.
In contrast to the situation at higher temperatures, the dilute
$e^+e^-$ plasma at lower temperatures is inefficient in shielding
isolated charges. Charge neutrality then requires a proton and an
electron to move together and the effective proton diffusion
constant is the smaller of the electron diffusion constant and
the proton diffusion constant.
Nuclei have to drag along a cloud of
$Z_i$ electrons. Naively, we may expect the diffusion constant for
nuclei to be suppressed by a factor of $1/Z_i$ relative to that of
the proton-electron system. This is because there are $Z_i$ as many
electrons scattering of photons and the probability of a change in
momentum of the nucleus-electron-cloud system is $Z_i$ times
as large as that for the proton-electron system.
Below $T\approx 0.1$ keV proton and nuclei diffusion is again limited by
Coulomb scattering as a result of the temperature
dependent increase of the
Coulomb cross section.

It is instructive to compute the diffusion length of baryons,
$d(t)$. This is the root-mean-square distance a baryon diffuses in time $t$.
The diffusion length $d_{100}=d/R$ in the early universe
becomes (Applegate et al. 1987)
$$d_{100}(t)=\biggl({6\over R^2}\int_0^tD(t^{\prime})dt^{\prime}\biggr
)^{1\over 2}\ ,\eqno(40)$$
and a simple integration through the $e^+e^-$-annihilation epoch
yields the results displayed in Figure 12.
In this figure we show $d_{100}^n$ for neutrons
as a function of the temperature
of the universe
for three
different local baryon-to-photon ratios
($\eta=1\times 10^{-4}$, $\eta=3.5\times 10^{-10}$,
and $\eta=3.5\times
10^{-12}$). This figure also shows $d_{100}^p$ for protons.
It is evident that baryon diffusion is inhibited in
large amplitude fluctuations.

An approximate time scale to double the size of a
square wave fluctuation by baryon diffusion is
$$\tau_b\approx {L^2\over D_b}\ ,\eqno(41)$$
where $D_b$ is the relevant baryon diffusion constant and $L$ is the
fluctuation length scale. Note that the damping time scale
is independent of the fluctuation amplitude $\Delta$.
In Figure 13 we compare the time scales for neutron diffusion
limited by magnetic moment scattering (short-dashed), neutron diffusion
limited by nuclear scattering in a fluctuation with
$\eta=6\times 10^{-9}$ (dashed-dotted), and proton diffusion
(long-dashed) to the Hubble time (solid) between temperatures of
$T=10$ MeV and $T=10$ keV.
For a fluctuation length of $L_{100}$ $^>_{\sim}$ 1 m we find proton diffusion
to
have negligible effects above $T\approx 20$ keV. Neutron diffusion is
seen to be dependent on the local proton density. A high density
fluctuation with characteristic interior
baryon-to-photon ratio $\eta\approx 10^{-4}$ and
fluctuation length $L_{100}\approx 1$ m
will not be much affected by neutron diffusion
whenever the temperature is above $T\approx 50$ keV.

\vskip 0.15in
{\bf\centerline {3. Evolution of Entropy Fluctuations
in the Early Universe}}
{\bf\centerline {- a Numerical Simulation}}
\vskip 0.1in

We have simulated the evolution of high-amplitude
baryon-to-entropy fluctuations between temperatures of $T=100$ MeV
and $T=10$ keV. The numerical simulation presented here does not
include the synthesis of elements below temperatures
of $T\approx 100$ keV. We evolve a spherical region of the universe,
which contains an initially Gaussian shape fluctuation at its center.
We assume that the horizon volume is filled with such spherical regions
so that the mean separation of fluctuation centers is
approximatly twice the cell
radius. These assumptions lead to reflective boundary conditions for
baryon number flux and heat flux at the spherical cell's edge.
We represent a fluctuation by a Lagrangian grid with sixty zones
in the spherical cell.

Baryon diffusion, neutrino inflation and photon inflation are
treated in a simple intuitive manner. Initially, we take each zone
to be in pressure equilibrium.
This pressure equilibrium configuration is characterized
by a temperature deviation
$\delta^0$ and proton (neutron) fluctuation amplitudes
$\Delta_p^0$ ($\Delta_n^0$) in each zone.
Diffusive processes during a time interval $dt$ will change
the temperature differences in each zone by
$\delta^0\mapsto\delta^1$, and
the baryon densities by $\Delta_{p,n}^0\mapsto\Delta_{p,n}^1$.
This will perturb the existing pressure equality, and will result in
an adiabatic expansion
(contraction) of zones, until pressure equilibrium
is re-established for new zone values of $\delta$, $\Delta_p$,
and $\Delta_n$. The simple algorithm used here has the advantage of
being easily tested against our analytic calculations for a two-zone
square wave fluctuation.

To obtain the changes in energy density from diffusive processes in zone $i$
during a time interval $dt$, we rewrite the heat diffusion
equation (14) in integral form, which yields
$$\delta_i^1=\delta_i^0+{dt\over (x_i^3-x_{i-1}^3)}
\Biggl[6{D\over R^2}x_i^2{(\delta_{i+1}^0-\delta_i^0)\over
(x_{i+1}-x_{i-1})}-3\int_0^{x_{i-1}}x^2{d\delta\over dt}dx\Biggr]
\ ,\eqno(42)$$
where $x_i$ is the outer boundary of the $i$-th zone.
In this expression $x_i$ is measured on our comoving $T=100$ MeV
scale, so that $x_i$ is short hand notation for $x_{100}^i$.
For
homogeneous neutrino heat transport we find a change in the
energy density in time interval $dt$ given by
$$\delta_i^1=\delta_i^0\biggl(1-dtF^{\star}{g_{\nu}
\over g_{eff}}
{1\over l_{\nu}}\biggr)\ .\eqno(43)$$
For this special case the heating is uniform across all zones $i$ in the
fluctuation.

Changes in baryon density due to baryon diffusion are obtained
from an integral diffusion equation similiar to equation (42).
However, efficient neutron diffusion at low temperatures necessitates
the use of an implicit algorithm. We replaced neutron
densities at time $t_0$ on the right-hand side of equation (42)
by neutron densities at time $t_1$.
The resulting matrix equation can be solved by the standard
methods of Gaussian elimination and back-substitution.
In addition we added weak interaction conversion
between neutrons and protons.
Weak reaction rates, as well as the evolution of scale factor
$R$, temperature $T$, and statistical weight $g$ are treated as in
the primordial nucleosynthesis code of
Wagoner, Fowler, \& Hoyle (1967) as updated by Kawano (1992).

Having computed the changes in energy and baryon content
of the fluctuation in time
interval $dt$, from heat transport, baryon diffusion
and weak interactions, we then allow each zone to expand
adiabatically to a new pressure equilibrium.
For this purpose we fix the common pressure in all zones
after adiabatic expansion
to equal the average pressure in the whole spherical cell in which the
simulation is performed.
Thus, it is easy to compute the fractional volume change
of zone $i$, $(dV/V)_i$, needed to attain pressure equilibrium
$$\biggl({dV\over V}\biggr)_i\approx 3(\delta_i^1-\delta_i^0)\ .
\eqno(44)$$
We can leave out changes in the baryon density in equation (44) since
the associated changes in baryon pressure in different zones are
negligible compared to the pressure changes from heat transport.
This approximation is good since the baryon-to-photon ratio is
everywhere small in our calculation.

Both photons and neutrinos eventually will decouple on the scale of the
fluctuation
The assumption of pressure
equilibrium breaks down when the photons decouple from the fluctuation
($l_{\gamma}>L$). The simulation of the hydrodynamic
expansion of a fluctuation is nontrivial since the photon
mean free path $l_{\gamma}$ varies strongly over the spherical cell
radius.
We encounter situations where photons are still in the optically thick
regime in high-density regions of the fluctuation but are in the optically
thin regime in lower density regions. In this case the damping of the
fluctuation proceeds by diffusive heat transport in the
core region of a fluctuation where the baryon density is high,
and by hydrodynamic expansion
near the boundary of the fluctuation. Numerical simulation of this process
requires the introduction of independent sets of photon zones
and baryon zones.

Neutrinos are easier to treat.
The neutrino mean free path $l_{\nu}$ is independent of
the baryon overdensity $\Delta$ and neutrinos
go quickly from the optically thick limit to the optically
thin limit. Therefore it is a fairly accurate approximation to
switch from diffusive neutrino
heat transport to homogeneous neutrino heat transport when the
neutrino
mean free path becomes larger than the width $a$ of the Gaussian
fluctuation.
\vskip 0.15in

In the following we will summarize the results of our numerical
simulation. Figures 14-16 show the evolution of three sample
high-density spherically condensed Gaussian entropy fluctuations
for temperatures between $T=100$ MeV and $T=10$ keV. The initially
Gaussian fluctuations are characterized by four quantities,
the average baryon-to-photon ratio
$\eta$ (equivalently $\Omega_b h^2$) within the
spherical simulation volume, the radius of the spherical simulation
volume $L_{100}^s$, the ratio of the radius $L_{100}^s$
to the Gaussian
width of the baryon density distribution $a$ ,
and the ratio of the baryon density at
the core of the fluctuation $n_b^H$ to the baryon density at the
edge of the simulation volume $n_b^L$, $\Lambda=(n_b^H/n_b^L)$.
For each sample we show nine evolution snapshots of the
fluctuation at different temperatures.
We plot the logarithm of the neutron and proton fluctuation amplitudes
($log_{10}\Delta_n$ and $log_{10}\Delta_p$)
as functions of the radius
$r_{100}=r/R$.
We reference all proper lengths to the values they would
have at $T=100$ MeV to remove the effects of the overall Hubble expansion.
Solid lines indicate the local proton fluctuation amplitude,
$\Delta_p$,
whereas dotted lines indicate the local neutron fluctuation
amplitude, $\Delta_n$.

In Figure 14 the effects of neutrino inflation are evident. Very
high-amplitude fluctuations with an initial
central baryon-to-entropy ratio
$(n_b/s)_i$ $ ^>_{\sim}$ $1.5\times 10^{-5}$ are damped to a characteristic
flat-top
fluctuation with final baryon-to-entropy ratio
$(n_b/s)_f\approx 1.5\times 10^{-5}$
(corresponding to an ultimate baryon-to-photon ratio of
$\eta_f\approx 10^{-4}$).
The limiting amplitude is attained between temperatures
of $T\approx 100$ MeV and $T\approx 20$ MeV depending on initial fluctuation
parameters.

The modifications of fluctuations
due to neutron diffusion are evident in all three sample evolutionary plots.
Figure 16 illustrates that small-scale fluctuations, $L$
$^<_{\sim}$ 0.5 m,
are almost completely homogenized and damped out before the onset of
primordial nucleosynthesis at $T\approx 100$ keV.

Photon inflation and the hydrodynamic expansion of fluctuations
at low temperatures are seen to be rapid and rather violent events.
This is evident in
Figure 15. Between temperatures of $T\approx 35$ keV and $T\approx 10$
keV these processes lower the
core proton density of fluctuations
by several orders of magnitude.
The effects of proton diffusion at low temperatures ($T\approx 20$ keV)
add to the
violent decay of such small-scale fluctuations.

\vskip 0.15in
{\bf\centerline {4. Conclusions}}
\vskip 0.1in

In the present study we have investigated the evolution of
fluctuations in the baryon-to-photon ratio
(entropy fluctuations) between the end of a cosmic electroweak
phase transition at $T\approx 100$ GeV and the end of primordial
nucleosynthesis at $T\approx 5$ keV. Our study focused on
nonlinear sub-horizon scale fluctuations.
Entropy fluctuations may result from an epoch of the early universe
where there is a departure from local thermodynamic equilibrium
such as a first order electroweak phase transition.
We have shown that entropy fluctuations evolve rapidly to an isobaric
character.

Fluctuations are found to damp by five different
physical processes: neutrino inflation in the diffusive and
homogeneous limits; photon inflation and hydrodynamic expansion;
and baryon diffusion.

Neutrino inflation provides the dominant damping of entropy fluctuations
between temperatures of $T\approx 100$ GeV and $T\approx 1$ MeV.
Fluctuations are damped by neutrino heat transport in such
a way that almost any initial fluctuation amplitude converges to
a generic final fluctuation amplitude.
Fluctuations of high initial amplitude
and with an initial fluctuation length
$L_{100}$ $^<_{\sim}$ $10^{-14}$ m
become damped to a characteristic baryon-to-entropy ratio
$(n_b/s)\approx 2\times 10^{-8}$. These fluctuations evolve
to this amplitude between $T\approx 100$ GeV and
$T\approx 1$ GeV, depending on their initial characteristics.
Fluctuations in the length scale regime
$10^{-5}$ m $ ^<_{\sim}\ L_{100}$ $^<_{\sim}\ 10^{-1}$ m
are damped to a
characteristic baryon-to-entropy ratio $(n_b/s)\approx
1.1\times 10^{-5}$ (corresponding to an ultimate baryon-to-photon
ratio $\eta\approx 8\times 10^{-5}$)
at an epoch between $T\approx 100$ MeV and $T\approx 1$ MeV.

The effects of photon inflation and hydrodynamic expansion on entropy
fluctuations are shown to be significant at temperatures
below $T\approx 30$ keV, corresponding to the approximate completion of
$e^+e^-$-annhilation. Fluctuations with an initial fluctuation
length
$L_{100}$ $^<_{\sim}\ 1$ m are almost
completely erased by these processes by the time the temperature
has reached $T\approx 20$ keV.

Our numerical calculations of fluctuation evolution include all
relevant diffusive and hydrodynamic damping processes.
We do not include any modifications due to nuclear reactions here.
Clearly our results have important implications for primordial
nucleosynthesis yields in inhomogeneous cosmologies. We will
address these issues in future work.

\vskip 0.15in
The authors wish to thank the Institute
of Geophysics and Planetary Physics at Lawrence Livermore National
Laboratory, and acknowledge useful discussions with
C.R. Alcock, G.J. Mathews, and B.S. Meyer.
We also wish to thank L.H. Kawano.
This work was supported by NSF grant PHY-9121623 and IGPP grant 93-22.

\vfil\eject

{\bf\centerline {References}}
\vskip 0.15in

\noindent Alcock, C. R., Fuller, G. M., \& Mathews, G. J. 1987,
ApJ, 320, 439

\noindent
Alcock, C. R., Dearborn, D. S., Fuller, G. M., Mathews, G. J., \& Meyer,
B. S. 1990, Phys.
\indent
Rev. Lett., 64, 2607

\noindent
Applegate, J. H., \& Hogan, C. J. 1985, Phys. Rev., D31, 3037

\noindent
Applegate, J. H., Hogan, C. J., \& Scherrer, R. J. 1987,
Phys. Rev., D35, 1151

\noindent
Applegate, J. H. 1991, Nucl. Phys., A527, 195c

\noindent
Banerjee, B., \& Chitre, S. M. 1991, Phys. Lett., B258, 247

\noindent
Brown, F. R. et al. 1990, Phys. Rev. Lett., 65, 2491

\noindent
Cohen, A., Kaplan, D., \& Nelson, A. 1990, Phys. Lett., B245, 561

\noindent
Cohen, A., Kaplan, D., \& Nelson, A. 1991a, Nucl. Phys., B349, 727

\noindent
Cohen, A., Kaplan, D., \& Nelson, A. 1991b, Phys. Lett., B263, 86

\noindent
Dine, M., Huet, P., Singleton, R., \& Susskind, L. 1991, Phys. Lett.,
B257, 351

\noindent
Dolgov, A., \& Silk, J. 1992, BERKELEY-CfPA-TH-92-04, preprint

\noindent
Fuller, G. M., Mathews, G. J., \& Alcock, C. R. 1988, Phys. Rev.,
D37, 1380

\noindent
Fuller, G. M., Jedamzik, K., Mathews, G. J., \& Olinto, A. 1993,
Phys. Rev. D,

\indent
submitted

\noindent
Gould, R. J. 1993, ApJ, in press

\noindent
Hagedorn, R. 1971, in {\it Thermodynamics of Strong Interactions},
CERN Lecture Notes,
\indent
Cern 71-12

\noindent
Heckler, A., \& Hogan, C. J. 1993, Phys. Rev. D, submitted

\noindent
Hogan, C. J. 1978, MNRAS, 185, 889

\noindent
Hogan, C. J. 1993, ApJ Lett., submitted

\noindent
Jedamzik, K., Fuller, G. M., \& Mathews, G. J. 1993, ApJ, in press

\noindent
Kajantie, K., \& Kurki-Suonio, H. 1986, Phys. Rev., D34, 1719

\noindent
Kawano, L. H. 1992, FERMILAB-PUB-92/04-A, preprint

\noindent
Kirzhnits, D. A., \& Linde, A. D. 1976, Ann. Phys. (N.Y.), 101, 195

\noindent
Kolb, E. W., Turner, M. S. 1989, in {\it The Early Universe},
Addison-Wesley

\noindent
Kurki-Suonio, H. 1988, Phys. Rev., D37, 2104

\noindent
Kurki-Suonio, H., Matzner, R. A., Centrella, J., Rothman, T.,
\& Wilson, J. R. 1988, Phys.

\indent
Rev., D38, 1091

\noindent
Malaney, R. A., \& Butler, M. N. 1989, Phys. Rev. Lett., 62, 117

\noindent
Malaney, R. A., \& Mathews, G. J. 1993, Phys. Rep., in press

\noindent
Mathews, G. J., Meyer, B. S., Alcock, C. R., \& Fuller, G. M. 1990,
ApJ, 358, 36

\noindent
McLerran, L. 1989, Phys. Rev. Lett., 62, 1075

\noindent
McLerran, L., Shaposhnikov, M., Turok, N., \& Voloshin, M. 1991,
Phys. Lett., B256, 451

\noindent
Misner, C. W. 1967, Nature, 214, 40

\noindent
Mueller, B. 1985, in {\it The Physics of the Quark-Gluon Plasma},
Springer-Verlag

\noindent
Nelson, A. 1990, Phys. Lett., B240, 179

\noindent
Peebles, P. J. E. 1965, ApJ, 142, 1317

\noindent
Peebles, P. J. E. 1971, in {\it Physical Cosmology}, Princeton
University Press

\noindent
Petersson, B. 1991, Nucl. Phys., A525, 237c

\noindent
Shaposhnikov M. E. 1986, JETP Lett., 44, 465

\noindent
Shaposhnikov M. E. 1987, Nucl. Phys., B287, 757

\noindent
Shaposhnikov M. E. 1988, Nucl. Phys., B299, 797

\noindent
Terasawa, N., \& Sato, K. 1989a, Prog. Theor. Phys., 81, 254

\noindent
Terasawa, N., \& Sato, K. 1989b, Phys. Rev., D39, 2893

\noindent
Terasawa, N., \& Sato, K. 1989c, Prog. Theor. Phys., 81, 1085

\noindent
Terasawa, N., \& Sato, K. 1990, ApJ Lett., 362, L47

\noindent
Turok, N., \& Zadrozny, P. 1990, Phys. Rev. Lett., 65, 2331

\noindent
Turok, N., \& Zadrozny, P. 1991, Nucl. Phys., B358, 471

\noindent
Wagoner, R. V., Fowler, W. A., \& Hoyle, F. 1967, ApJ, 148, 3

\noindent
Witten, E. 1984, Phys. Rev., D30, 272

\vfil\eject

{\bf\centerline {Figure Captions}}

\vskip 0.1in
\noindent
{\bf Figure 1:}
We show a schematic representation of the ratio of entropy
density in fluctuations to the average entropy density plotted
against the ratio of the baryon number density in fluctuations to
the average baryon number density. We show adiabatic, isothermal,
isocurvature, and isobaric (heavy line) fluctuations.
The dotted lines represent the tracks for the nearly adiabatic hydrodynamic
expansion/contraction of fluctuations to pressure equilibrium.

\vskip 0.1in
\noindent
{\bf Figure 2:} Three square wave fluctuations are shown. The net
baryon number density $n_b(x)$ and the plasma temperature $T(x)$
is shown as a function of length scale $x$. The size of the high density
region of a fluctuation, or fluctuation length scale, is $L$, and the mean
separation between centers of fluctuations is $L^s$.

\vskip 0.1in
\noindent
{\bf Figure 3:} The product of cosmic scale factor and temperature
$R(T)(T/100 {\rm MeV})$ plotted against temperature
$T$ in MeV. We take $R=1$ at $T=100$ MeV.

\vskip 0.1in
\noindent
{\bf Figure 4:} The proportionality constant $f_p$ (solid line)
between pressure
exerted by relativistic particles and the energy density in
relativistic particles plotted against temperature $T$ in MeV.
Also shown is the temperature dependence of the proportionality
constant
between the deviation in relativistic energy density and the
baryon overdensity within a fluctuation (dashed line).

\vskip 0.1in
\noindent
{\bf Figure 5:}  We plot the ratio of the temperature (or energy density)
deviations in a fluctuation for two different baryon equation of state
against temperature. Here $\delta_{pg}$ is the deviation using a perfect
gas Maxwell-Boltzmann equation of state for neutrons and protons,
while $\delta$ is the deviation including baryon-antibaryon pairs.
We show this ratio for three baryon overdensities $\Delta$.

\vskip 0.1in
\noindent
{\bf Figure 6:} The proper radius in units of $10^{-8}$ m multiplied by
$R^{-1}$ is shown as a function of temperature for three different
initial fluctuation amplitudes. In all three fluctuations an initial
radius $L_{100}=10^{-8}$ m is assumed.
The fluctuations
are distinguished by different initial baryon-to-entropy ratios
$(n_b/s)_i=10^{-1}$, $10^{-3}$, and $10^{-5}$. The
neutrino mean free path $l_{\nu}^{100}$ is also displayed
(dashed line).

\vskip 0.1in
\noindent
{\bf Figure 7:} Displayed are the effects of neutrino inflation
between the epochs of $T=100$ GeV and $T=100$ MeV on entropy fluctuations.
We give the final baryon-to-entropy ratio $(n_b/s)_f$ of neutrino-inflated
two-zone fluctuations as a function of the initial fluctuation radius
$L^i_{100}$ in meters. We show five different initial
($T=100$ GeV) values of $(n_b/s)_i$.
Dashed lines indicate that $(n_b/s)_f$ depends on uncertainties
in the QCD equation of state in the strong coupling limit. The approximate
horizon scale at $T=100$ GeV (denoted electroweak horizon EWH) is on the
far right hand side of the horizontal scale.

\vskip 0.1in
\noindent
{\bf Figure 8:} Effects of neutrino inflation between the epochs $T=100$ MeV
and $T=1$ MeV. Notation is as in Figure 7. We give $(n_b/s)_f$ for
fluctuations with initial length scale $L^i_{100}$ in meters and initial
($T=100$ MeV) amplitude $(n_b/s)_i$ as labeled.
The approximate horizon scale at $T=100$ MeV (QCDH) is shown.
The dashed line divides fluctuations containing more (right
side) and less (left side) baryonic mass than the baryon mass in the
QCD-horizon.

\vskip 0.1in
\noindent
{\bf Figure 9:}
Effects of neutrino inflation on entropy fluctuations
over the history of the universe
between epochs of
$T=100$ GeV and $T=1$ MeV.
Notation is as in Figures 7 and 8.

\vskip 0.1in
\noindent
{\bf Figure 10:} Photon mean free path $l_{\gamma}^{100}=l_{\gamma}/R$ in
meters as a function of temperature.
Results are displayed for three different proton fluctuation amplitudes
$\Delta_p$. A fraction of the closure density
in baryons $\Omega_b=0.013h^{-2}$ has been
assumed.

\vskip 0.1in
\noindent
{\bf Figure 11:} The fluctuation damping rates due to
diffusive photon inflation
(dashed) and hydrodynamic expansion against Thomson drag
(dashed-dotted) and
the Hubble expansion rate (solid) are given as functions of
temperature.
The fluctuation length is taken as $L_{100}=1$ m.

\vskip 0.1in
\noindent
{\bf Figure 12:} The proton- (dashed)
and neutron- (solid) diffusion lengths $d_{100}=d/R$
in meters given as functions of temperature. We assume that the
diffusive random walk starts at $T=100$ MeV.
The neutron
diffusion length is shown for three different
baryon-to-photon ratios $\eta$.

\vskip 0.1in
\noindent
{\bf Figure 13:} The inverse of fluctuation damping time scales
from baryon diffusion are displayed as functions of temperature.
Shown are the neutron diffusion damping rate limited by
$ne$-scattering
(short-dashed), neutron diffusion damping rate limited by $np$-scattering
(dashed-dotted), and proton diffusion damping rate
limited by $e\gamma$-scattering
(long-dashed). For comparision the Hubble expansion rate (solid)
is given.
The key in the upper left-hand corner of the figure
displays the dependence of the different time scales on
assumed fluctuation radius $L_{100}=1$ m, average
fraction of closure density in baryons
$\Omega_b=0.25h^{-2}$, and proton fluctuation amplitude
$\Delta_p=1$.

\vskip 0.1in
\noindent
{\bf Figure 14:} The evolution of a sample entropy
fluctuation over the history of the universe between epochs
$T=100$ MeV and $T=30$ keV. We show nine evolution
snapshots at different temperatures $T$. Each evolution snapshot
displays the logarithm of the proton fluctuation amplitude
$log_{10}\Delta_p$ (solid) and the
logarithm of the neutron fluctuation amplitude
$log_{10}\Delta_n$ (dotted) as functions of radial
coordinate $r_{100}=r/R$ in meters.
The assumed initial fluctuation characteristics are
displayed at the bottom of the plots.

\vskip 0.1in
\noindent
{\bf Figure 15:} Same as Figure 14 but with different initial
fluctuation characteristics as shown.

\vskip 0.1in
\noindent
{\bf Figure 16:} Same as Figure 14 but with different initial
fluctuation characteristics as shown.

\end